\documentclass[twocolumn,trackchanges]{aastex63}
\usepackage{graphicx}
\usepackage{amsmath}
\usepackage{url}
\usepackage{comment}
\usepackage{ulem}
\usepackage{color}
\usepackage[left]{lineno}
\def\dotdeg{\hbox{\,$.\!\!^{\circ}$}}

\newcommand\kms{\hbox{km$\,$s$^{-1}$}}

\newcommand\VLSR{\hbox{$V_{\rm LSR}$}}

\received{2022 August 18}
\revised{2022 November 23}
\accepted{2022 November 25}

\submitjournal{ApJ}

\shorttitle{Discovery of the ``Tadpole'' Molecular Cloud}
\shortauthors{Kaneko et al.}

\begin{document}
\title{Discovery of the Tadpole Molecular Cloud near the Galactic Nucleus}

\correspondingauthor{Miyuki Kaneko}
\email{miyukikaneko@keio.jp}

\author[0000-0003-4732-8196]{Miyuki Kaneko}
\affil{School of Fundamental Science and Technology, Graduate School of Science and Technology, Keio University, 3-14-1 Hiyoshi, Kohoku-ku, Yokohama, Kanagawa 223-8522, Japan}

\author[0000-0002-5566-0634]{Tomoharu Oka}
\affiliation{School of Fundamental Science and Technology, Graduate School of Science and Technology, Keio University, 3-14-1 Hiyoshi, Kohoku-ku, Yokohama, Kanagawa 223-8522, Japan}
\affiliation{Department of Physics, Institute of Science and Technology, Keio University, 3-14-1 Hiyoshi, Kohoku-ku, Yokohama, Kanagawa 223-8522, Japan}

\author[0000-0003-3853-1686]{Hiroki Yokozuka}
\affiliation{Department of Physics, Institute of Science and Technology, Keio University, 3-14-1 Hiyoshi, Kohoku-ku, Yokohama, Kanagawa 223-8522, Japan}

\author[0000-0003-2735-3239]{Rei Enokiya}
\affiliation{Department of Physics, Institute of Science and Technology, Keio University, 3-14-1 Hiyoshi, Kohoku-ku, Yokohama, Kanagawa 223-8522, Japan}

\author[0000-0001-8147-6817]{Shunya Takekawa}
\affiliation{Faculty of Engineering, Kanagawa University 3-27-1 Rokkakubashi, Kanagawa-ku, Yokohama, Kanagawa 221-8686. Japan}

\author[0000-0002-9255-4742]{Yuhei Iwata}
\affiliation{Division of Science, National Astronomical Observatory of Japan, 2-21-1 Osawa, Mitaka, Tokyo 181-8588, Japan}
\affiliation{Center for Astronomy, Ibaraki University, 2-1-1 Bunkyo, Mito, Ibaraki 310-8512, Japan}

\author[0000-0002-1663-9103]{Shiho Tsujimoto}
\affiliation{School of Fundamental Science and Technology, Graduate School of Science and Technology, Keio University, 3-14-1 Hiyoshi, Kohoku-ku, Yokohama, Kanagawa 223-8522, Japan}

\begin{abstract}
   In this paper, we report the discovery of an isolated, peculiar compact cloud with a steep velocity gradient at $2\farcm 6$ northwest of Sgr A*. This ``Tadpole'' molecular cloud is unique owing to its characteristic head-tail structure in the position-velocity space.  By tracing the CO {\it J}=3--2 intensity peak in each velocity channel, we noticed that the kinematics of the Tadpole can be well reproduced by a Keplerian motion around a point-like object with a mass of $1\!\times\! 10^{5}\,M_{\odot}$. Changes in line intensity ratios along the orbit are consistent with the Keplerian orbit model. The spatial compactness of the Tadpole and absence of bright counterparts in other wavelengths indicate that the object could be an intermediate-mass black hole.
\end{abstract}
\keywords{galaxies: nuclei --- Galaxy: center --- ISM: clouds --- ISM: molecules}

\section{Introduction}
\label{sec:intro}
It is widely accepted that large galaxies host a central supermassive black hole (SMBH) with millions to billions times the mass of the Sun (e.g., \citealt{Kormendy95}; \citealt{Kormendy13}). A potential scenario for SMBH formation is based on intermediate-mass black holes (IMBHs), which have masses of $10^{2\mbox{--}5}\, M_{\odot}$ (e.g., \citealt{Mezcua17}).  Thus, detecting and studying IMBHs in detail are essential for understanding the formation and evolution of galactic nuclei.
Numerous IMBH candidates have been identified in centers of globular clusters (e.g., \citealt{Kiziltan17}), in nuclei of dwarf galaxies (e.g., \citealt{Reines13, Baldassare15}), or as ultraluminous X-ray sources in extragalaxies (e.g., \citealt{Farrell09}).

\begin{figure*}[htbp]
   \centering
   \includegraphics[scale=0.9]{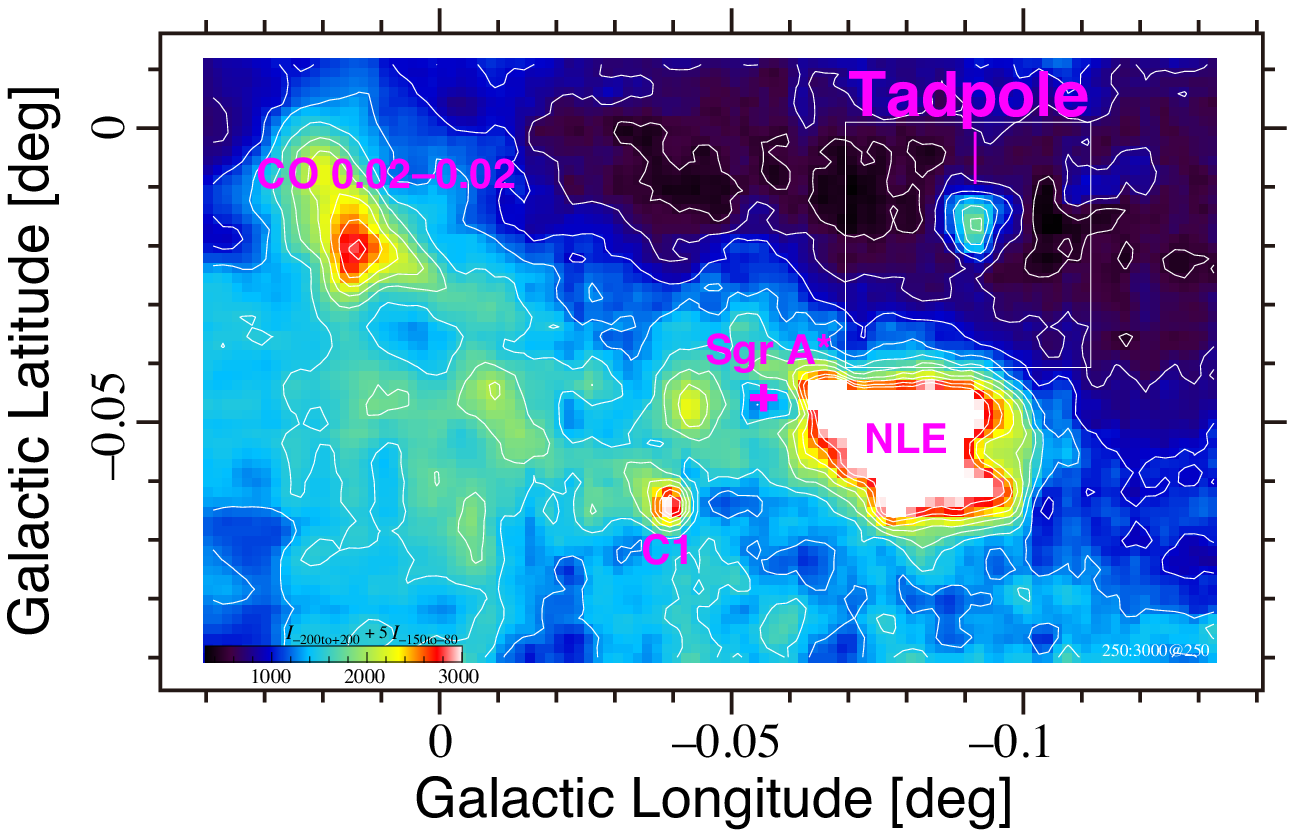}
   \caption{Map of velocity-integrated CO {\it J}=3--2 emission. White contours are drawn at 250 K \kms\ intervals from 250 K \kms. The integrated intensity was calculated by $\int_{-200}^{200} T_{\rm MB} dV\! +\! 5 \int_{-150}^{-80} T_{\rm MB} dV$ to emphasize the Tadpole, which appears at $(l, b)\!=\! (-0\fdg 090, -0\fdg 014)$. CO 0.02--0.02 (\citealt{Oka99, Oka08}), the C1 cloud (\citealt{Oka11,Takekawa17}), and the negative longitude extension of the circumnuclear disk (NLE; \citealt{Oka11,Takekawa17}) also appear in this map. The white rectangle indicates the area presented in Figures \ref{fig:lvmap}--\ref{fig:counterpart}}
   \label{fig:guidemap}
\end{figure*}

In the central molecular zone (CMZ) of our Galaxy, a number of compact ($d\! <\! 5$ pc) clouds with extraordinary broad velocity width ($\Delta V\! >\! 50$ \kms) have been detected (e.g., \citealt{Oka98, Oka99, Oka12, Oka22}). These peculiar clouds, namely, high velocity-dispersion compact clouds (HVCCs), have been assumed to be accelerated by supernova explosions, protostellar outflows, and/or cloud-to-cloud collisions (e.g., \citealt{Oka22}).  Subsequently, it was determined that the kinematics of CO--0.40--0.22, which is one of the most energetic HVCCs, can be well reproduced by a cloud being gravitationally kicked by a point-like mass of $\sim\! 10^{5}$ $M_{\odot}$ (\citealt{Oka16, Oka17}). The absence of any bright object near the point-like mass suggest that it may be an IMBH.  Subsequently, it was also suggested that HVCCs HCN--0.009--0.044 (\citealt{Takekawa19a}), HCN--0.085--0.094 (\citealt{Takekawa20}), and CO--0.31+0.11 (\citealt{Takekawa19b}) were  driven by an IMBH. These discoveries yielded a new method of finding non-luminous massive objects, such as inactive and wandering BHs.  To date, five IMBH candidates, including IRS13E (\citealt{Tsuboi19}), have been reported in the Galactic CMZ.

When searching for gravitationally kicked gas in the CMZ, we noticed an isolated HVCC in the CO {\it J}=3--2 data obtained with the James Clerk Maxwell Telescope (JCMT; \citealt{Parsons18, Eden20}). It appears as an isolated compact cloud at $(l, b)\!\simeq\! (-0\fdg 090, -0\fdg 014)$, which corresponds to $\sim\! 2\farcm 6$ Galactic northwest of Sgr A* (Figure \ref{fig:guidemap}), with LSR velocities between $-140$ \kms\ and $-90$ \kms. It stands out with its peculiar appearance and very high CO {\it J}=3--2/CO {\it J}=1--0 intensity ratio ($R_{3\mbox{--}2/1\mbox{--}0}\!=\! 1.8$  \citealt{Oka22}), which exceeds double of the CMZ average ($R_{3\mbox{--}2/1\mbox{--}0}\!\sim\! 0.7$  \citealt{Oka07, Oka12}). In this paper, we report the discovery of the so-called ``Tadpole'' molecular cloud, which is listed as id.75 in the catalog of \citet{Oka22}.  Throughout this paper, the distance to the Galactic center is assumed to be $D\! =\! 8.3$ kpc (\citealt{Gravity18}).

\begin{figure*}[htbp]
   \centering
   \includegraphics[scale=0.5]{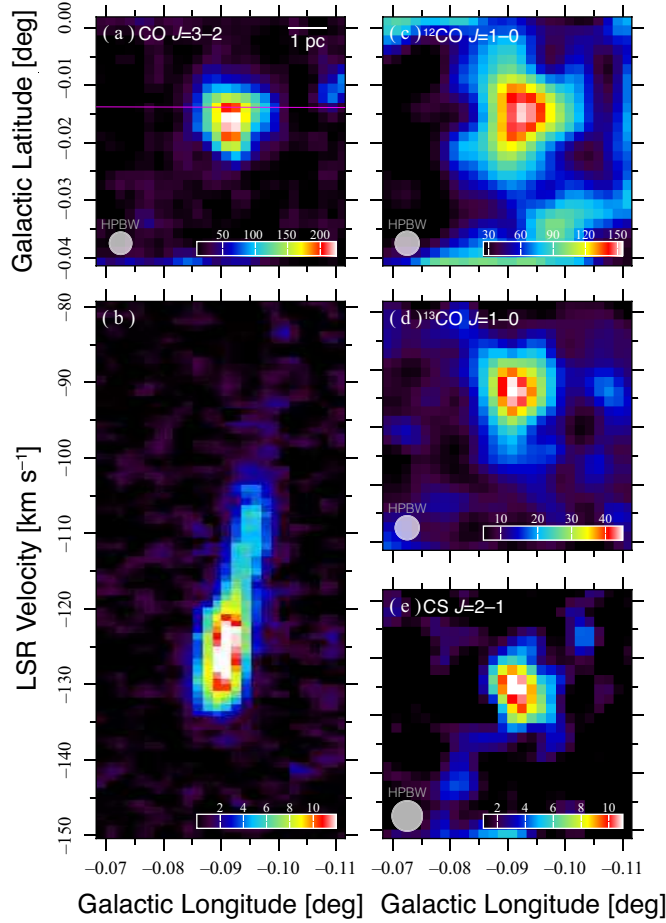}
   \caption{(a) Map of velocity-integrated CO {\it J}=3--2 emission.  The velocity range for integration is $V_{\rm LSR}\!=\! -140$ to $-80$ \kms. The intensity unit is K \kms. (b) Longitude-velocity map of CO {\it J}=3--2 emission at $b\!=\! -0\fdg 014$ (the magenta line in panel (a)). (c) Map of velocity-integrated $^{12}$CO {\it J}=1--0 emission. (d) Map of velocity-integrated $^{13}$CO {\it J}=1--0 emission. (e) Map of velocity-integrated CS {\it J}=2--1 emission. The integration ranges for these lines are the same as that of the CO {\it J}=3--2 line.}
   \label{fig:lvmap}
\end{figure*}

\section{Data}
We first discovered the Tadpole in the JCMT CO {\it J}=3--2 data, and confirmed it in the CO {\it J}=1--0 and CS {\it J}=2--1 line data obtained with the Nobeyama Radio Observatory (NRO) 45 m telescope. These data sets are briefly described below.

\subsection{CO {\it J}=3--2 Line}
The $^{12}$CO {\it J}=3--2 line ($345.795990$ GHz) observations of the CMZ were performed using the JCMT from 2013 July to 2014 July (\citealt{Parsons18}). The Heterodyne Array Receiver Program (HARP; \citealt{Buckle09}) and autocorrelation spectral imaging system (ACSIS) were used during these observations. The half-power beam width (HPBW) of the telescope was approximately $14\arcsec$ at 345 GHz. The ACSIS was operated in the $1$ GHz bandwidth ($976.56$ kHz resolution) mode. During these observations, the system noise temperature ($T_{\rm sys}$) ranged between $100\mbox{--}200$ K. The rms noise level of the image cubes was between $0.4$ K and $0.84$ K. The details of the CO {\it J}=3--2 data and JCMT observations are presented in \citet{Parsons18}. We use the data after resampling onto a $7\farcs 5\!\times\! 7\farcs 5\!\times\! 1$ \kms\ grid.

\subsection{CO {\it J}=1--0 Lines}
The CO {\it J}=1--0 observations of the CMZ were performed using the Nobeyama Radio Observatory (NRO) 45 m telescope (\citealt{Tokuyama19}).  The $^{12}$CO {\it J}=1--0 (115.27120 GHz) line data were obtained from 2011 January 19 to 29 using the 25 beam array receiver system (BEARS; \citealt{Sunada00}). As the receiver backend, the AC45 spectrometer system (\citealt{Sorai00}) was employed in the 500 MHz bandwidth (0.5 MHz resolution) mode. The $^{13}$CO {\it J}=1--0 (110.20135 GHz) line data were obtained from 2016 February to March using the four-beam receiver system on the 45 m telescope (FOREST; \citealt{Minamidani16}). The spectral analysis machine on the 45 m telescope (SAM45; \citealt{Kuno11, Kamazaki12}) was operated in the 1 GHz (244.14 kHz resolution) mode. The HPBW of the telescope was approximately $15\arcsec$ at 115 and 110 GHz.  The typical $T_{\rm sys}$ was $\sim\! 800$ K and $150\mbox{--}300$ K during the $^{12}$CO and $^{13}$CO observations, respectively.  The data were resampled onto a $7\farcs 5\!\times\! 7\farcs 5\!\times\! 2$ \kms\ grid.  The rms noise levels of the resultant $^{12}$CO and $^{13}$CO data cubes were 1.0 K and 0.2 K in main-beam temperature ($T_{\rm MB}$), respectively.

\subsection{CS {\it J}=2--1 Line}
The CS {\it J}=2--1 (97.98096 GHz) line observations of the CMZ were performed during the NRO 45 m Telescope Large Program through 2019 January--May and 2020 January--April. The mapping area was set to $-1\fdg 5\!\le\! l\!\le\!+1\fdg 5$ and $-0\fdg 25\!\le\! b\!\le\! +0\fdg 25$. The FOREST receiver and SAM45 spectrometer were used. The SAM45 was operated in the 1 GHz bandwidth ($244.14$ kHz resolution) mode.  The HPBW of the telescope was $\simeq\! 19\arcsec$ at 86 GHz. The $T_{\rm sys}$ ranged from 150--300 K during the CS {\it J}=2--1 line observations. The data were resampled onto a $7\farcs 5\!\times\! 7\farcs 5\!\times\! 2$ \kms\ grid. The rms noise level of the resultant CS {\it J}=2--1 data cube was $0.14$ K in $T_{\rm MB}$. The details of the CS data and NRO 45 m observations will be presented in the forthcoming paper (Takekawa et al. 2022, in preparation).

\begin{figure}[htbp]
   \centering
   \includegraphics[scale=0.4]{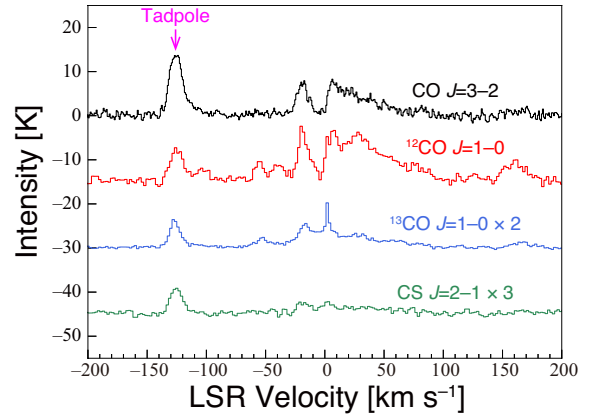}
   \caption{Observed line spectra at (--0\dotdeg091, --0\dotdeg016) at where the CO {\it J}=3--2 profile of the Tadpole is widest. Black, red, blue and green lines show CO {\it J}=3--2, $^{12}$CO {\it J}=1--0, $^{13}$CO {\it J}=1--0 and CS {\it J}=2--1 line spectra, respectively. }
   \label{fig:spectrum}
\end{figure}

\begin{figure*}[htbp]
   \centering
   \includegraphics[scale=0.5]{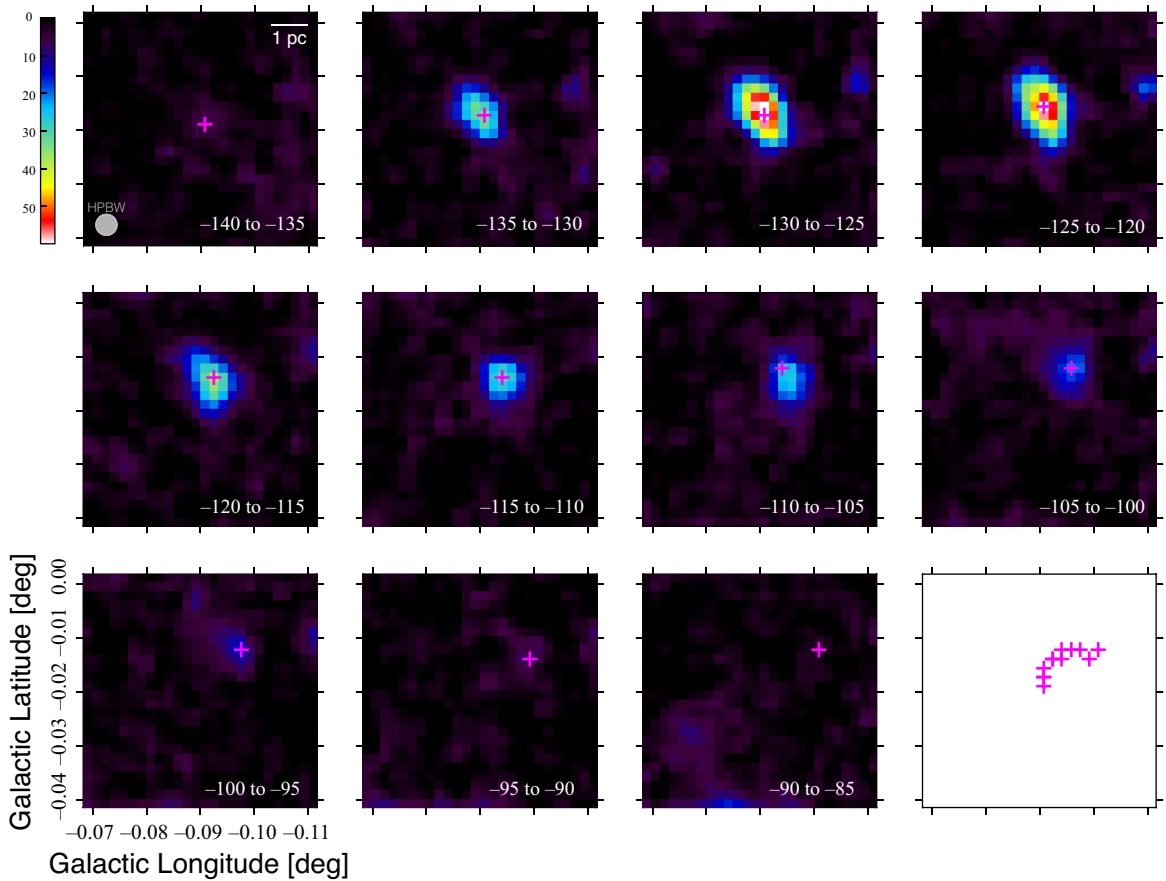}
   \caption{Velocity channel maps of the CO {\it J}=3--2 line from $\VLSR\! =\! -140$ to $-85$ \kms\ near the Tadpole. The angular area is the same as that depicted in Figure \ref{fig:lvmap}a, c--e. The intensity unit is K \kms. The velocity range for integration is indicated at the bottom right corner of each panel. The magenta cross denotes the position of the maximum intensity pixel in each panel. The distribution of the maximum intensity pixels is shown in the bottom right panel.}
   \label{fig:channelmap}
\end{figure*}

\section{Results}
\subsection{Spatial and Velocity Structure} \label{subsec:structure}
Figure \ref{fig:lvmap} shows the spatial and velocity structure of the Tadpole in various molecular lines. The Tadpole appears as a compact, well-defined clump in velocity-integrated maps (Figure \ref{fig:lvmap}a, c--e). In the CO {\it J}=3--2 integrated map (Figure \ref{fig:lvmap}a), the full width of half maximum (FWHM) angular diameter of the Tadpole is $28\arcsec$. This angular diameter is $2$ times the angular resolution ($14\arcsec$), corresponding to a distance of $1.1$ pc from the Galactic center. In the $^{13}$CO {\it J}=1--0 and CS {\it J}=2--1 maps, the Tadpole has angular sizes similar to that depicted in the CO {\it J}=3--2 map, while the $^{12}$CO {\it J}=1--0 map depicts an appearance that is larger by a factor of 2 compared with those depicted in the other maps.

The velocity extent of the Tadpole is from $\VLSR\!\simeq \! -135$ \kms\ to $-90$ \kms\ (Figure \ref{fig:lvmap}b), indicating that it may be in the CMZ. The longitude-velocity behavior of the Tadpole is characterized by a ``head-tail" structure, from which its name is derived. It depicts a steep velocity gradient of $\left| \Delta V/\Delta l \right|\!\sim\! 16$ \kms pc$^{-1}$. The tail is obscure in the $^{13}$CO {\it J}=1--0 and CS {\it J}=2--1 data sets.  We show the spectra of CO and CS lines in Figure \ref{fig:spectrum}.

We learned that CO {\it J}=3--2 line can trace the kinematics of the Tadpole best because of its high intensity and well-defined appearance in the {\it l--b--V} space. Figure \ref{fig:channelmap} shows the velocity channel maps of CO {\it J}=3--2 emission. The Tadpole appears with an elliptical shape in each velocity channel, changing its angular size with respect to velocity. It has the largest angular size of $36\arcsec\!\times\! 24\arcsec$ in the $\VLSR\! =\! -130$ to $-125$ \kms\ channel, while the smallest size of $\sim\! 20\arcsec$ is detected at both velocity ends. These angular sizes, which are comparable to the telescope's HPBWs, indicate that the spatial structure of the Tadpole is not well resolved using these single-dish observations; thus, the actual angular size must be far smaller than those observed.

The CO {\it J}=3--2 line emission shows a  maximum intensity of $64$ K \kms\ in the $-130$ to $-125$ \kms\ channel. We identified that the intensity maximum pixel in the Tadpole at each velocity channel continuously changes with respect to velocity. Note that these intensity maximum pixels trace an arc in the plane of the sky (Figure \ref{fig:channelmap}, bottom right panel). This behavior will be analyzed in more detail in the discussion section (\S\ref{sec:kinematics}).

\begin{figure}[htbp]
   \centering
   \includegraphics[scale=0.65]{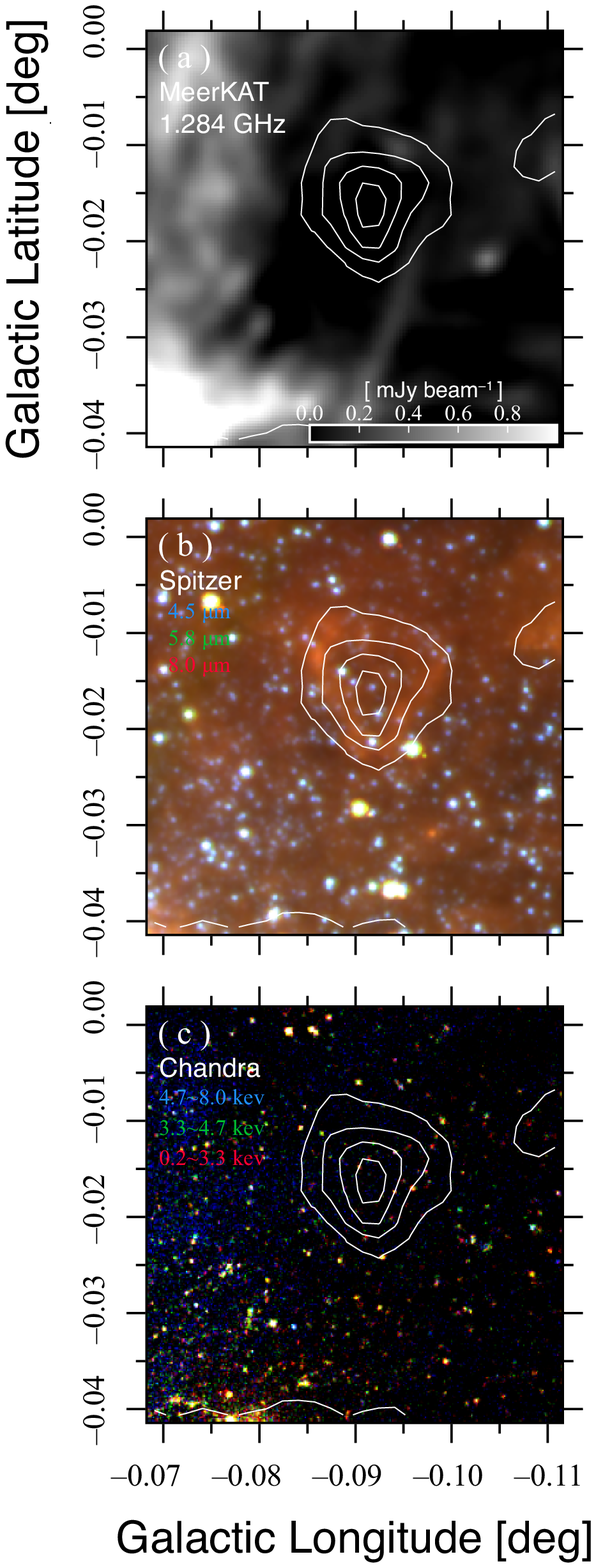}
   \caption{Multiwavelength view near the Tadpole. White contours show the CO {\it J}=3--2 integrated intensity with a $50$ K \kms\ interval.
      (a) Radio continuum image at 1.284 GHz obtained with the MeerKAT. (b) Composite mid-infrared image obtained with the Spitzer Space Telescope.  The image depicts 8.0 $\mu$m band flux in red, 5.8 $\mu$m in green, and 4.5 $\mu$m in blue. (c) Composite X-ray image obtained with the Chandra X-ray Observatory. The image depicts $0.2\mbox{--}3.3$ keV count rate in red, $3.3\mbox{--}4.7$ keV in green, and $4.7\mbox{--}8.0$ keV in blue.}
   \label{fig:counterpart}
\end{figure}

\subsection{Physical Parameters}\label{subsec:physicalParameters}
The physical size of the Tadpole was evaluated as the size parameter; defined by $S\!\equiv\!D\tan\left(\sqrt{\sigma_{l}\sigma_{b}}\right)$ to be $2.2$ pc. The velocity dispersion was calculated to be $\sigma_{V}\! =\! 22$ \kms. These give a size-linewidth coefficient ($\sigma_{V}/S^{0.5}$) of $15$ \kms pc$^{-0.5}$ and virial theorem mass ($M_{\rm VT}\!\equiv\! 8.7 S\sigma_{V}^2/G$) of $4.7\!\times\! 10^{5}$ $M_{\odot}$. The molecular gas mass ($M_{\rm gas}$) was estimated by summing the CO {\it J}=3--2 line integrated intensity and using the CO(3--2)-to-H$_2$ conversion factor ($X_{\rm CO 3\mbox{--}2} \!=\! 1.4\!\times \! 10^{20}$ [cm$^{-2}$ (K \kms )$^{-1}$] \citealt{Oka22}) to be $6.6\!\times\! 10^{2}$ $M_{\odot}$. The $X_{\rm CO 3\mbox{--}2}$ we employed is close to $1\!\times \! 10^{20}$ that estimated from an LVG calculation for $n({\rm H}_2)\!=\!1\!\times\!10^4$ cm$^{-3}$, $T_{\rm k}\!=\!60$ K, and $N({\rm CO})/dV\!=\!1\!\times\!10^{17}$ cm$^{-2}$ (\kms)$^{-1}$.  Using the physical parameters described above, we derived the dynamical timescale ($t_{\rm dyn}\!\equiv\! S/\sigma_{V}$), kinetic energy ($E_{\rm kin}\!\equiv\! 1.5 M_{\rm gas}\sigma_{V}^{2}$), and kinetic power ($P_{\rm kin}\!\equiv\! E_{\rm kin}/t_{\rm dyn}$) of the Tadpoe as $2.2\!\times\!10^{4}$ yr, $9.4\!\times\! 10^{48}$ erg, and $1.4\!\times\! 10^{38}$ erg s$^{-1}$, respectively.  The kinetic power of the Tadpole is equal to $3.7 \times 10^4 \, L_{\odot}$ which is far greater than those of molecular outflows from massive YSOs (0.01--100 $L_{\odot}$; \citealt{Maud15}).

The virial theorem mass and molecular gas mass of the Tadpole yield a significantly high virial parameter, $M_{\rm VT}/M_{\rm gas}\!\sim\! 700$. This indicates that the Tadpole is not in gravitational equilibrium.  Following the method described in \citet{Stark89}, we estimated the self-gravity of the Tadpole and tidal force by the Galactic potential. The self-gravity of the cloud estimated here ($\sim\! 10^{-10}$ m s$^{-2}$) is 2.5 orders of magnitude weaker than the tidal force ($\sim\! 4\!\times\!10^{-8}$ m s$^{-2}$). These assessments clearly demonstrate that the Tadpole cannot be bound by its self-gravity. This strongly indicates to the presence of an object with a mass comparable to the virial mass ($4.7\!\times\! 10^{5}$ $M_{\odot}$) inside the Tadpole.

\subsection{Multiwavelength View}\label{sec:multiwavelength}
To search for the driving source behind the Tadpole, we referred to the 1.284 GHz data obtained using MeerKAT (\citealt{Heywood22}), mid-infrared data obtained using the Spitzer Space Telescope (\citealt{Ramirez08, Churchwell09}), and X-ray data obtained at the Chandra X-ray Observatory (\citealt{Muno09}).

No radio source brighter than 0.4 mJy beam$^{-1}$ at 1.284 GHz was detected near the angular extent of the Tadpole, while a faint filament was seen in the Galactic south (Figure \ref{fig:counterpart}a). In the mid-infrared image (Figure \ref{fig:counterpart}b), we identified a faint triangular rim which may be the irradiated surface of the Tadpole. Considering the angular resolution of the Spitzer ($\sim\!2\arcsec$), the mid-infrared rim defines the angular extent of the Tadpole better than the CO {\it J}=3--2 appearance. We also checked the ATLASGAL (870$\mu$m) and Hi-GAL (70, 160, 250, 350 and 500 $\mu$m) images (\citealt{Contreras13, Molinari16}), and found no separate feature toward the Tadpole above the sea of intense dust emission. A bright point-like source near the Galactic southwestern edge of the Tadpole is the long-period variable star V4872 Sgr (\citealt{Matsunaga09}). In addition, we also identify dozens of point-like, mid-infrared sources toward the Tadpole. The X-ray image also shows numerous faint point-like sources toward the Tadpole (Figure \ref{fig:counterpart}c). Although the nature of these mid-infrared/X-ray point-like sources is unclear, we will refer to them in the discussion section (\S\ref{sec:origin}). In short, the multiwavelength view confirms the absence of any energetic objects that can drive the Tadpole.

\section{Discussion}
\subsection{Tracing Molecular Gas Kinematics}\label{sec:kinematics}
The continuous change in the CO {\it J}=3--2 intensity maximum pixel along the arc with velocity (\S\ref{subsec:structure}) suggests that the bulk of warm molecular gas belongs to a certain trajectory, such as a closed orbit. To determine the orbit trajectory accurately, we constructed a CO {\it J}=3--2 data cube with a $2\arcsec\!\times\! 2\arcsec\!\times\! 1$ \kms\ grid. Then, we calculated the accurate intensity peak position as the CO emission center of gravity using $5\!\times\!5$ pixels around the intensity maximum pixel in the Tadpole for each $1$ \kms\ width velocity channel.

\begin{figure}[htbp]
   \centering
   \includegraphics[scale=0.55]{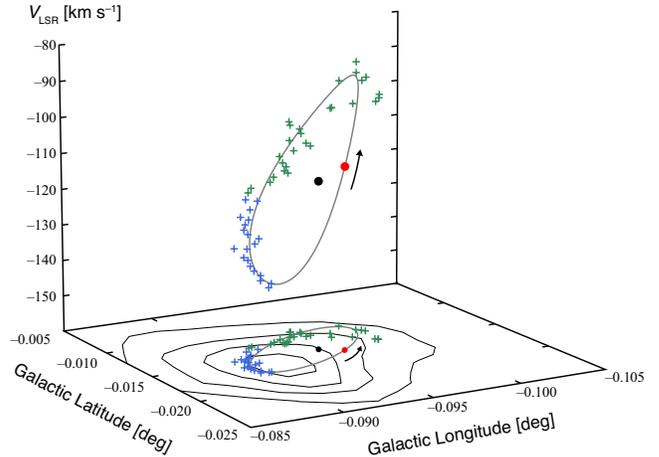}
   \caption{The {\it l--b--V} distribution of CO {\it J}=3--2 intensity peak positions in the Tadpole in each $1$ \kms\ width velocity channel. The head and tail of the Tadpole are denoted by blue and green, respectively. The {\it l--b} distribution is also shown in the base plane with CO {\it J}=3--2 contours. Grey lines represent the loci of the best-fit Keplerian orbit (Table \ref{table:parameters}). The black and red filled circles denote the positions of the dynamical center and pericenter, respectively. The black arrow indicates the direction of rotation.}
   \label{fig:kep3D}
\end{figure}

Figure \ref{fig:kep3D} shows the {\it l--b--V} and {\it l--b} distributions of the intensity peak position in the Tadpole for each velocity channel. The arc-shape apparent in the {\it l--b} distribution became clearer than that depicted in Figure \ref{fig:channelmap}. The {\it l--b--V} distribution demonstrates the striking continuity of peak positions in the velocity direction, indicating that the bulk of warm molecular gas in the Tadpole may predominantly follow a certain closed orbit.

\begin{deluxetable}{ll}
   \tablecaption{Parameters of the best-fit Keplerian Orbit}\label{table:parameters}
   \tablehead{
      \colhead{Parameter} & \colhead{Value}
   }
   \startdata
   Mass ($M_{\mathrm{dyn}}$) & $(1.01\pm 0.05)\times 10^{5}$  $M_{\odot}$ \\
   Semimajor axis ($a$) & $0.75 \pm 0.01$  $\mathrm{pc}$ \\
   Eccentricity ($e$) & $0.35 \pm 0.03$ \\
   Inclination ($i$) & $112\arcdeg \pm 1\arcdeg$   \\
   Argument of pericenter ($\omega$) & $82\arcdeg \pm 5\arcdeg$  \\
   Longitude of ascending node ($\Omega$) & $-35\arcdeg \pm 1\arcdeg$  \\
   Longitude offset ($l_{0}$) & $-0\fdg 0946 \pm 0\fdg 0001$  \\
   Latitude offset ($b_{0}$) & $-0\fdg 0146 \pm 0\fdg 0001$  \\
   Velocity offset ($V_{0}$) & $-114 \pm 1$  \kms \\
   \enddata
\end{deluxetable}

\subsection{Keplerian Orbit Fitting}\label{sec:fitting}
The arc-shaped spatial structure and continuous velocity change along the arc suggest that a rotational motion exists. Thus, we expect that the {\it l--b--V} behavior of warm molecular gas in the Tadpole can be reproduced by a Keplerian orbit around a point mass\footnote{Here we do not consider parabollic or hyperbolic orbits, although the orbit could not be an elliptical.}. We performed the fitting of a Keplerian orbit to the intensity peak positions (Figure \ref{fig:kep3D}) following the method described in \citet{Zhao09}.  We determined nine parameters. Five of these parameters are three-dimensional orbital parameters; namely, the semimajor axis $(a)$, eccentricity $(e)$, longitude of ascending node $(\Omega)$, argument of pericenter $(\omega)$, and inclination angle $(i)$. The remaining parameters are the mass $(M_{\rm dyn})$, line-of-sight velocity $(V_{\rm dyn})$ and {\it l--b} position $(l_{0},\,b_{0})$ of the dynamical center.
The distance to the Tadpole is assumed to be same as that to the Galactic center. The $\chi^2$ minimization approach was utilized in the {\it l--b--V} space. The $\chi^2$ consists of two terms, namely, the spatial ($\chi^2_{\rm s}$) and velocity terms ($\chi^2_{\rm v}$). The spatial term is defined by the sum of the orthogonal distances between the peak position and the modeled orbit divided by the square of the positional uncertainty, which was set to $0.15$ pc ($3\farcs 6$). The velocity term is, similarly, the sum of the velocity deviations between those of the peak positions and nearest positions in the modeled orbit divided by a square of the velocity uncertainty, which was set to $1$ \kms.

The best-fit parameters with $1\sigma$ uncertainties are listed in Table \ref{table:parameters}. The {\it l--b--V} locus of the orbit is presented in Figure \ref{fig:kep3D}. Note that the best-fit solution is bivalent because the orbits with $i\!\rightarrow\! (180\arcdeg\! -i)$ and $\Omega\!\rightarrow\! (\Omega +180\arcdeg)$ yield the same {\it l--b--V} locus. The fitting result indicates  the presence of a point mass of $1.0\!\times\!10^5$ $M_{\odot}$ in the northwestern periphery of the Tadpole (Figure \ref{fig:kep3D}).

\begin{figure}[htbp]
   \centering
   \includegraphics[scale=0.38]{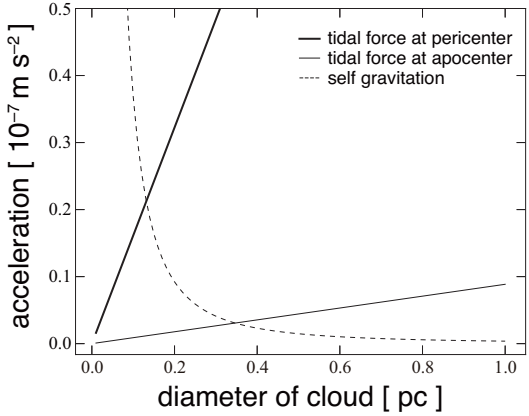}
   \caption{Plots of the tidal force due to the point mass (solid line) and self-gravity of the cloud (dashed lines) working on the head versus the head diameter. A central mass of $10^5$ $M_{\odot}$ and head mass of $660$ $M_{\odot}$ were assumed. The thick solid line indicates self-gravity at the pericenter distance (0.49 pc) while the thin solid line indicates that at the apocenter (1.01 pc). }
   \label{fig:tidal}
\end{figure}

\subsection{Tidal Stability of the Head}
We examined the tidal stability of the head, assuming the Newtonian potential of a point mass, following the method described in \citet{Stark89}.
The tidal force caused by the point mass and self-gravity doing work on the head were calculated as functions of the head diameter (Figure \ref{fig:tidal}). A central mass of $1\!\times\! 10^5$ $M_{\odot}$ and head mass of $660$ $M_{\odot}$ were assumed. The curves of the two forces at the pericenter of the best-fit orbit (0.5 pc) intersect at $d\!=\! 0.14$ pc, while those at the apocenter intersect at $d\!=\!0.35$ pc. Because the observed head diameter ($\sim\! 1$ pc) is larger than the critical value (0.14 pc) by a factor of seven, the Tadpole head is tidally unstable at the pericenter. If the head includes a self-gravitating core smaller than 0.14 pc, it may be able to survive several turns. It is possibly that the Tadpole has been trapped by the gravitational potential of the $10^5$ $M_{\odot}$ point mass, being stretched by the strong tidal force to form the characteristic head-tail structure. This situation is similar to the simulation of a cloud tidally disrupted by a super massive black hole (\citealt{Saitoh14}).  Note that the Tadpole must have been lost its angular momentum to be captured during the encounter with the point mass.

\begin{figure}[htbp]
   \centering
   \includegraphics[scale=0.5]{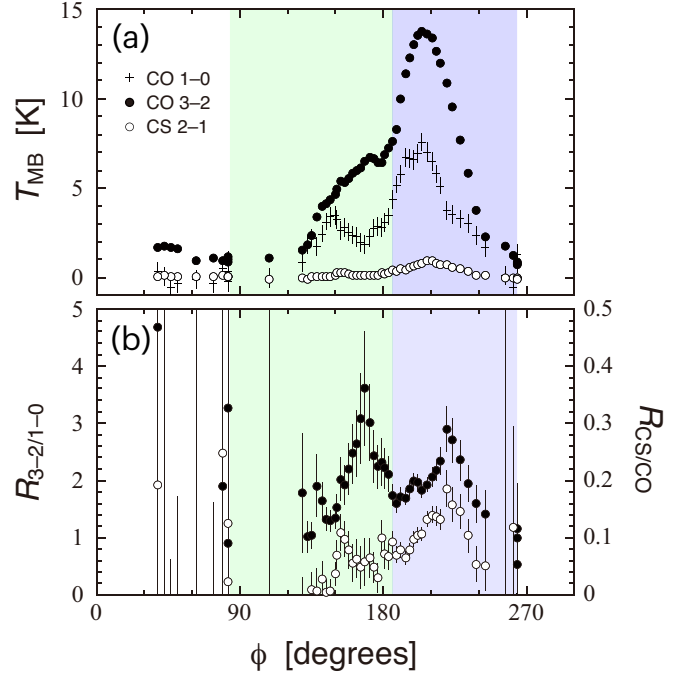}
   \caption{(a) Plots of CO {\it J}=3--2 (filled circles), CO {\it J}=1--0 (crosses), and CS {\it J}=2--1 (open circles) main-beam temperatures at the CO {\it J}=3--2 intensity peaks versus orbital phase. (b) Plots of $R_{3\mbox{--}2/1\mbox{--}0}$ (filled circles) and $R_{\rm CS/CO}$ (open circles) at the CO {\it J}=3--2 intensity peaks versus orbital phase. The phase ranges of the head and tail of the Tadpole are depicted in blue and green shades, respectively.}
   \label{fig:ratio}
\end{figure}

\subsection{Intensity Ratios along the Orbit}\label{sec:dis_ratio}
Here, we refer to the CO {\it J}=1--0, CO {\it J}=3--2, and CS {\it J}=2--1 line intensities, as well as CO {\it J}=3--2/CO {\it J}=1--0 ($R_{3\mbox{--}2/1\mbox{--}0}$) and CS {\it J}=2--1/CO {\it J}=1--0 ($R_{\rm CS/CO}$) line intensity ratios to examine the Keplerian orbit scenario. The critical densities of the CO {\it J}=1--0, CO {\it J}=3--2, and CS {\it J}=2--1 transitions are approximately $10^{2.5}$, $10^{4}$, and $10^{5}$ cm$^{-3}$, respectively, while their upper state energies are $5.6$, $33$, and $7.1$ K, respectively. Thus, $R_{\rm CS/CO}$ is sensitive to the variations in density while $R_{3\mbox{--}2/1\mbox{--}0}$ is affected by both the temperature and density.  Figure \ref{fig:ratio} shows plots of line intensities and intensity ratios along the best-fit orbit. The motion along the orbit increases the orbital phase ($\phi$) at all times, where the pericenter of the orbit corresponds to $\phi\!=\! 0\arcdeg$.

Both ratios exhibit higher values at the head, indicating that both the density and temperature are enhanced there. At the beginning of the tail, $\phi\!\sim\! 170\arcdeg $, $R_{3\mbox{--}2/1\mbox{--}0}$ demonstrates a prominent peak while $R_{\rm CS/CO}$ does not. This suggest that the temperature is enhanced at the beginning of the tail.  Then, both ratios decrease toward the tip of the tail. The higher temperature in the head may have been caused by the shock occurring at the first encounter with the point mass. This shock may be caused by collisions of gas clumps at the pericenter, where adjacent orbits are very close. In a fluid mechanical treatment, the same process is described as the ``tidal compression"\citep{Saitoh14}.  The gas accumulation near the apocenter and/or the presence on a self-gravitating core can cause the higher density in the head. The tail may be really the low density, tidally stretched tail of the head. The temperature increase at the beginning of the tail could be caused by another shock at the congestion near the apocenter. Thus, the behaviors of intensity ratios are consistent with the Keplerian orbit scenario for the Tadpole.

\begin{figure}[htbp]
   \centering
   \includegraphics[scale=0.55]{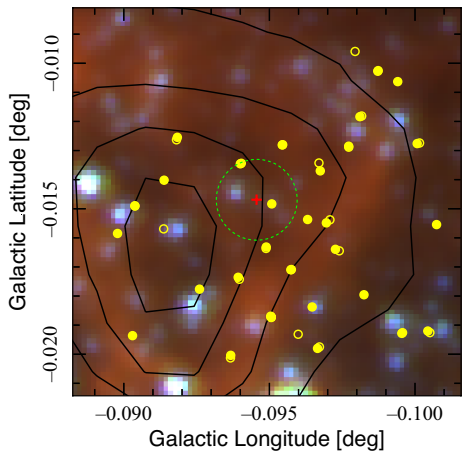}
   \caption{Distribution of X-ray point-like sources (open yellow circles: \citealt{Muno09}, filled yellow circles: \citealt{Zhu18}) superimposed on the composite mid-infrared image (color; same as Fig.\ref{fig:counterpart}b). The red cross denotes the dynamical center of the best-fit Keplerian orbit, and the green dotted circle has a $5\arcsec$ radius. Black contours highlight the CO {\it J}=3--2 integrated intensity with a 50 K \kms\ interval.
   }
   \label{fig:X-ray}
\end{figure}

\begin{deluxetable*}{cccccccc}
   \tablecolumns{8}
   \tablecaption{Properties of X-ray and mid-infrared sources near the point mass}
   \label{table:counterpart}
   \tablehead{
      &\colhead{$l$}\vspace{-0.4cm}& \colhead{$b$} & \colhead{$F_{{\rm 2\mbox{--}10 keV}}$$^{\text{a}}$}& \colhead{[3.6 $\mu$m]$^{\text{b}}$ } & \colhead{[4.5 $\mu$m]$^{\text{b}}$}& \colhead{[5.8 $\mu$m]$^{\text{b}}$}& \colhead{[8.0 $\mu$m]$^{\text{b}}$}\\
      \colhead{ID}\vspace{-0.4cm}&&&&&&&\\
      & \colhead{(deg)} & \colhead{(deg)} & \colhead{(10$^{-18} \, \mathrm{W / m^{2}}$)} & \colhead{(mag)} & \colhead{(mag)} & \colhead{(mag)} & \colhead{(mag)}
   }
   \startdata
   SSTGC 489898$^{\text{b}}$&--0.09513 & --0.01486 & 1.66 & 10.284 & \nodata  & \nodata  &  \nodata \\
   SSTGC 490125$^{\text{b}}$&--0.09384 & --0.01444 &  \nodata & 9.154 & 9.003 & 8.422 & 8.217 \\
   CXOGC 174526.9--290124$^{\text{c}}$&--0.09400 & --0.01346 & 16.72 & \nodata  &  \nodata &  \nodata &  \nodata \\
   \enddata
   \tablecomments{\\
      $^{\mathrm{a}}$ The X-ray (2--10 keV) energy flux quoted from \citet{Zhu18}.\\
      $^{\mathrm{b}}$ The source IDs and mid-infrared magnitudes are quoted from \citet{Ramirez08}.\\
      $^{\mathrm{c}}$ The source ID defined in \citet{Muno09}.
   }
\end{deluxetable*}

\subsection{What is a Point Mass?}\label{sec:origin}
The Keplerian orbit model requires the presence of a huge point mass at the dynamical center.  What is this point-like massive object?  The mass of $10^{5}\,M_{\odot}$ is larger than the Arches or Quintuplet clusters ($\sim\! 10^4\,M_{\odot}$ \citealt{Figer1999a, Figer1999b}).  According to the best-fit model, a mass of $10^{5}\,M_{\odot}$ must be concentrated within a radius significantly smaller than 0.5 pc (the pericenter distance), resulting in an average mass density higher than that of the Arches cluster ($\rho\!\sim\! 2\!\times\! 10^{5}\,M_{\odot}\,\mathrm{pc}^{-3}$ \citealt{Espinoza09}). Anyway, the absence of a bright infrared counterpart toward the Tadpole (\citealt{Ramirez08, Churchwell09, Molinari11}) clearly rules out a stellar cluster from being a candidate for the point mass.

The most promising candidate for the point-like massive object in the Tadpole may be an IMBH. We searched for a counterpart for this point mass referring to X-ray images (\S\ref{sec:multiwavelength}). Figure \ref{fig:X-ray} shows the positions of point-like sources in the X-ray images superimposed on the composite mid-infrared images around $(l, b)\!=\! (l_0, b_0)$. The small statistical uncertainties in the {\it l--b} position of the dynamical center (Table \ref{table:parameters}) should be considered with some caution because the entire gas mass in the Tadpole is not confined within the single closed orbit. Notice that three point-like sources in the X-ray and mid-infrared reside within a $5\arcsec$ ($\sim\! 10$ times of the positional ambiguity of the dynamical center) angular distance from $(l_0, b_0)$ (Table \ref{table:counterpart}).  One source, SSTGC 489898, is apparent in both X-ray and mid-infrared images. The X-ray detected sources, SSTGC 489898 and CXOGC 174526.9--290124, exhibit hard spectra, indicating that they may be at the same distance to the Galactic center. No time variation has been detected from these sources \citep{Zhu18}. If we assume the standard accretion disk model ($L\!=\!\dot{M} c^2 /12$), their X-ray luminosities correspond to $\dot{M}\!\sim\!1\!\times\!10^{-14}\;M_{\odot} \mathrm{yr}^{-1}$.  This very low mass accretion rate could be a challenge to our interpretation.  Anyway, we suppose these three point-like sources are candidates for the luminous counterpart of the point mass.  Although we currently have very little knowledge on these point-like sources, they should be considered as candidates for an IMBH in future studies.

\section{Conclusions}
We have discovered the so-called Tadpole, which is an isolated, peculiar compact cloud with an extraordinary velocity width and very high CO {\it J}=3--2/{\it J}=1--0 intensity ratio, at $2\farcm 6$ northwest of Sgr A*.   Our main conclusions can be summarized as follows:
\begin{enumerate}
   \item The Tadpole molecular cloud has a size of $\sim\! 1$ pc at the distance to the Galactic center and a velocity width of $\sim\! 50$ \kms .
   \item It demonstrates the characteristic ``head-tail" structure in position-velocity space, having a steep velocity gradient of $16$ \kms pc$^{-1}$.
   \item Its kinematics is well reproduced by a Keplerian motion around a point-like object with a mass of $1\!\times\! 10^{5}\,M_{\odot}$.
   \item It is plausible that the Tadpole has been trapped by the gravitational potential of the huge point mass, now being stretched by the strong tidal force.
   \item The behaviors of line intensity ratios ($R_{3\mbox{--}2/1\mbox{--}0}$ and $R_{\rm CS/CO}$) are consistent with the Keplerian orbit scenario.
   \item The absence of bright objects near the putative point-like object suggests that the object is an inactive intermediate-mass black hole (IMBH).
\end{enumerate}

These results are based on molecular line maps with $14\arcsec\mbox{--}19\arcsec$ resolutions obtained using single-dish telescopes. Future aperture synthesis observations of molecular lines with millimeter/submillimeter arrays will be able to delineate the orbit stream directly, increasing the reliability of the Keplerian orbit model for the Tadpole.
\\
\\

The results presented in this paper are based on data obtained using the Nobeyama Radio Observatory (NRO) 45-m telescope and James Clerk Maxwell Telescope (JCMT). The NRO 45-m radio telescope is operated by the Nobeyama Radio Observatory, a division of the National Astronomical Observatory of Japan.

The James Clerk Maxwell Telescope is operated by the East Asian Observatory on behalf of The National Astronomical Observatory of Japan, Academia Sinica Institute of Astronomy and Astrophysics, Korea Astronomy and Space Science Institute, National Astronomical Research Institute of Thailand, and Center for Astronomical Mega-Science (as well as the National Key R\& D Program of China with No. 2017YFA0402700). Additional funding support is provided by the Science and Technology Facilities Council of the United Kingdom, and participating universities and organizations in the United Kingdom and Canada.

We are grateful to the NRO staff and all members of the JCMT team for operating the telescope. T.O. acknowledges the financial support of JSPS Grant-in-Aid for Scientific Research (A) No. 20H00178.  S. Ta acknowledges support from JSPS Grant-in-Aid for Early-Career Scientists Grant Number JP19K14768.

\end{document}